\documentstyle[aps,prl,twocolumn,epsf,rotate]{revtex}

\begin{document}
\twocolumn[\hsize\textwidth\columnwidth\hsize\csname
@twocolumnfalse\endcsname

\title{Nature and number of distinct phases in the random field Ising model}

\draft

\author{Jairo Sinova$^{1,2,3}$, Geoff Canright$^{1,2,4}$} 
\address{$^{1}$Department of Physics,
University of Tennessee, Knoxville, Tennessee 37996}
\address{$^{2}$Department of Physics,
Indiana University, Bloomington, Indiana 47405-4201}
\address{$^{3}$Department of Physics,
The University of Texas, Austin, Texas 78712-1081}
\address{$^{4}$Physics Institute, University of Oslo, Norway}
\date{\today}
\maketitle

\begin{abstract}
We investigate the phase structure of the random-field Ising model
with a bimodal random field distribution.
Our aim is to test 
for the possibility of an equilibrium spin-glass phase, and for
replica symmetry breaking (RSB) within such a phase. We study a
low-temperature region where the spin-glass phase is thought to occur,  
but which has received little numerical study to date. We use the exchange
Monte-Carlo technique to acquire equilibrium information about the
model, in particular the $P(q)$ distribution
and the spectrum of eigenvalues of the spin-spin
correlation matrix (which tests for the presence of RSB). 
Our studies span the range in parameter space
from the ferromagnetic to the paramagnetic phase.
We find however no convincing evidence for
any equilibrium glass phase, with or without RSB, between these two
phases. Instead we find clear evidence (principally
from the $P(q)$ distribution) that there are only two phases
at this low temperature, with a discontinuity in the magnetization
at the transition like that seen at other temperatures.
\end{abstract}

\pacs{}

\vskip2pc]

\section{Introduction}

The random-field Ising model (RFIM) is one of the simplest problems
that one can define theoretically in which disorder is present and plays
a significant role in the physics. This problem gains further interest 
since a related random system which may be realized 
experimentally---namely, the site-diluted antiferromagnet in a uniform 
field---may be mapped onto the
RFIM. The RFIM has been subjected to considerable study, both
experimental and theoretical, for over 25 years \cite{recentreview}. 
Much effort has been focussed
on the nature of the phase diagram. It is known
rigorously that, for low temperatures and low random fields, the ferromagnetic
(FM) phase found in the pure system survives for $d \ge 3$ (where $d$
is the space dimension). However most other questions that one can ask
about this system for $d=3$ (hence beyond the range of validity of 
mean-field theory) remain unanswered. For example, is the phase transition
first-order or continuous? Is there a tricritical point, separating
a critical line from a line of first-order transitions? Does the answer to 
these questions, and/or the universality class of the critical line, 
depend on the distribution of random fields? And finally: is there a third,
spin-glass (SG) phase, in addition to (and likely separating) the FM and 
paramagnetic (PM) phases?

In this work we focus on this last question. Early numerical studies 
using mean-field theory \cite{Yoshizawa,Ro_Levin,Grest_Levin} 
defined a nonequilibrium ``domain state", characterized by various forms of 
irreversible behavior, of a similar nature to the irreversible 
behavior seen experimentally. This irreversibility region, found 
between the PM and FM phases, gives (at most) hints of the presence of an 
equilibrium glassy phase. Stronger arguments for a glassy phase were made
by de Almeida and Bruinsma \cite{Almeida}, who showed that,
although mean-field theory in strictly infinite dimension
gives replica symmetry \cite{Schneider}, in large dimensions there
is a stable glassy phase with replica symmetry breaking (RSB). 
This result is supported by the work of Mezard and Young
\cite{Mezard_and_Young} and of Mezard and Monasson
\cite{Mezard_and_Monasson}, who showed that there is a SG phase with
RSB in the limit of a large number of spin components, and by
Guagnelli {\it et al} \cite{Guagnelli_Marianari_Parisi}, who showed the
presence of multiple solutions to the mean-field equations in $d=3$,
at a temperature well above the FM ordering temperature. 
Subsequently, Brezin and de Dominicis \cite{Brezin} developed
a renormalization-group approach for a replicated theory,
and argued for RSB on the basis of a loss of stability of the 
fixed point. Finally, we mention exact numerical zero-temperature studies
\cite{Bastea} which show the appearance of extensive entropy
in the ground state at the point where FM order disappears.

Thus there are good reasons to consider the possibility of a SG
phase for the RFIM, located between the FM and PM phases, and 
likely (see \cite{Almeida}) more easily detected at low $T$ and large
random field. It is also of great interest to test for the presence
of RSB in this phase. To date, conclusive evidence for
RSB is confined to theoretical models in infinite dimension;
to confirm RSB in a three-dimensional problem---which
can be studied experimentally---would be an important step.

In earlier work \cite{scm,sccm} we have developed a method
for detecting RSB in finite-size numerical studies. The method
relies on examining the eigenvalues of the spin-spin correlation
matrix. We showed that, in the limit of large number of spins
$N$, there are multiple $O(N)$ eigenvalues of this matrix,
if \cite{sccm} and only if \cite{scm} there is RSB. This was 
clearly confirmed by Monte-Carlo (MC) studies of the infinite-ranged
Sherrington-Kirkpatrick problem, which is known to have RSB.
Unfortunately, this method gave no evidence for RSB
in a finite-(4)-dimensional version of the same problem.

We apply the same method to the 3-dimensional RFIM in this work.
We note that previous numerical studies \cite{forexample}
of phase transitions 
of the RFIM are almost entirely confined to two regions: either
relatively high temperatures and low random fields, or zero temperature.
However, based on the location of the irreversibility regions in phase space,
and on the results of Ref.\ \cite{Almeida}, one is most likely
to find a glassy phase by looking
in the high-field, low-$T$ region; this we do here.

Our results, in short, are as follows. We find some, rather ambiguous,
evidence for RSB in the behavior of the spin-spin eigenvalues,
over a small range of field, at fixed, low temperature. On either
side of this region we see clearly the PM and FM phases. Thus this
narrow region is either the sought-for glassy phase, or simply
the finite-size effects of the phase transition itself. Examination
of the order parameter function $P(q)$ enables us to rather conclusively
rule out the former explanation in favor of the latter: a FM/PM phase 
boundary. Finally, we see at the transition
the discontinuity (or near-discontinuity \cite{Falicov,Ogielski})
in magnetization that has been seen in many other studies, both at
zero \cite{Hartmann,Angles,Swift}
and at higher \cite{Machta,RiegerYoung,YoungNau} temperature. 

We organize the rest of 
this paper as follows. In Sec. II we briefly explain the general
technique of correlation matrix spectral analysis introduced in our previous
work \cite{scm}. In Sec. III we present the procedures and results of the
numerical analysis of the bi-modal RFIM model in a cross-section 
of the phase diagram. Finally in Sec. IV we present our conclusions.

\section{correlation matrix spectral analysis}

In considering frustrated and disordered systems such as spin glasses,
one is forced to broaden the concept of `ordering' in describing the
frozen phase. The key point, as recognized long ago by 
Edwards and Anderson  
\cite{EA}, is simply the freezing itself: long-time averages become nonzero.
One can detect this freezing through correlation functions:
if the freezing is long-ranged (i.e., truly a distinct phase),
then so are the correlations. Viewing the correlation function as a matrix,
one finds that ordering (freezing) appears in the spectrum of
the correlation matrix as an extensive (which we will also
call `large') eigenvalue \cite{Yang}.
Thus examination of this spectrum allows one to detect
ordering/freezing, while completely avoiding any need to
guess the {\it nature\/} of the ordering (i.e., the eigenfunction
corresponding to the large eigenvalue).

For Ising systems, a suitable correlation function is
\[
C_{ij}\equiv \langle S_i S_j \rangle\,\,;
\]
this is a real symmetric positive semi-definite matrix with
trace equal to the system volume $N$.
In the frozen phase at least one of the eigenvalues of $C_{ij}$
is extensive, i.e., of $O(N)$. For a simple Ising ferromagnet
without quenched disorder, freezing leads to the partitioning
of spin configuration space into two disjoint regions ($+$ and $-$);
each is frozen, and each has the same thermodynamic weight since they are
related by a symmetry of the Hamiltonian, namely spin inversion.
Since $C_{ij}$ is invariant under this symmetry, the two
(symmetry breaking) frozen states give only a single large eigenvalue.

In the case of replica symmetry breaking or RSB, there are multiple
frozen states which are not related by any symmetry of the Hamiltonian.
We have shown \cite{sccm} that, when there is RSB, there must
be more than one extensive eigenvalue in the spectrum of $C_{ij}$---assuming
of course that the thermal average is taken over the entire configuration 
space. We have also shown \cite{scm} that, in the absence of RSB, there can only
be one such extensive eigenvalue. Given these results, one can 
obtain a definitive answer to the question of whether or not a
system exhibits RSB, by determining whether there is or is not
more than one large eigenvalue in the spectrum.

This latter question is well defined, but is vulnerable to finite-size
uncertainties (as are other criteria such as the overlap distribution $P(q)$).
Here such uncertainties arise because the definition of an extensive
eigenvalue depends on taking the large-$N$ limit. Thus, in numerical
studies of finite systems, one must look for convincing evidence that 
one has indeed found the asymptotic behavior, at least of the two
largest eigenvalues of $C$, as $N$ increases. In particular, one must
determine if the second largest eigenvalue $\lambda_2$ grows
linearly with $N$ at large $N$. In the absence of RSB, $\lambda_2$
can grow \cite{scm,sccm} with $N$ as $\lambda_2$ $\sim N^{1-\delta}$ with
$\delta > 0$. RSB, in contrast, requires that $\delta$ be strictly zero.

Let us now discuss these ideas as applied specifically to the RFIM.
In the paramagnetic phase (for nonzero values of the random field)
there is freezing of the spins, which (on average) simply follow the 
random field. Thus the frozen configuration is unique, giving
a single large eigenvalue. In the ferromagnetic phase there is
a $+$ and a $-$ state. However, due to the random field, the two 
are no longer strictly related by symmetry; nor---due to 
a net field of $O(N^{1/2})$---do they have equal
thermodynamic weight. (These two states are termed `similar but
incongruent' by Huse and Fisher \cite{HuseFisher}.) Hence we again
expect a single extensive eigenvalue in the FM phase.

In this work we will look for signs of a spin-glass phase in a 
low-temperature region between the PM and FM phases. If the SG 
phase has RSB---and we are able to study a range of $N$ beyond the
threshold \cite{scm} for observing RSB---then we should find multiple 
large eigenvalues of $C$. Note that, if $\lambda_2/N$ decays over the
entire range of $N$ studied, and reaches a very small value within that range,
then that small value offers a rough upper bound for the thermodynamic
weight which may be assigned to any putative RSB state. Hence there may be RSB
in such a case; but it would be a `weak' RSB, where only one state 
has large thermodynamic weight. We also do not rule out the possibility
of a SG phase without RSB, i.e., with a single large eigenvalue.
In this case one needs other measures (such as qualitative
changes in the scaling of $\lambda_2/N$, in $P(q)$, or in the spin-glass
susceptibility $\chi_{SG}$) to distinguish the SG phase
from the other two.

\section{Numerical procedures and results}

\subsection{The numerical method}
The RFIM Hamiltonian is given by
\begin{equation}
{\cal H}=-J\sum_{\langle i j\rangle} S_i S_j -\sum_i h_i Si
\end{equation}
with $\langle i j\rangle$ indicating the sum over nearest neighbors, $J$ being
the strength of the ferromagnetic coupling, and $h_i$
being a quenched random variable. In this work we will take $J=1$,
so that temperatures and fields are measured in units of $J$. 
The two common probability distributions for
$h_i$ are a Gaussian distribution, with variance $\Delta$, or the bi-modal
distribution
\[
P(h_i)=\frac{1}{2}[\delta(\Delta-h_i)+\delta(\Delta+h_i)]\,\,.
\]
As we noted in the Introduction, there is some uncertainty as to which
features of the phase diagram are sensitive to the choice between these 
two distributions. For example, Aharony \cite{Aharony} predicts
a tricritical point for the bimodal distribution, but not for the
Gaussian. We will work with the bimodal distribution, principally because
it enables faster Monte-Carlo simulation. Also, we know of no reason to
expect that a SG phase is more or less likely to occur for one or
the other distribution.

As is the case in all glassy systems, the system sizes of 
Monte Carlo simulations
are severely restricted by long relaxation or equilibration times.
The typically long time that it takes to 
jump across large energy barriers in configuration space can be
greatly reduced by the parallel tempering or exchange Monte Carlo technique
\cite{Hukushima}. This technique consists of running simultaneously multiple 
replicas of the system at different temperatures, while allowing the swapping 
of spin configurations between adjacent temperatures. Such swaps are
attempted every $N_s$ Monte Carlo steps. By choosing the probability of swapping
correctly---meaning that the
probability of a given spin configuration being accepted from another 
replica at a different temperature is constrained to be consistent with 
the Boltzmann distribution at the original temperature---all copies of
the system remain in equilibrium under the swapping process.
This effectively enables the jumping of free-energy barriers,
since different replicas are likely to be in different local minima of 
the free energy.

Another important aspect of Monte Carlo simulations in quenched-disorder 
systems is a set of robust equilibration criteria,
needed to obtain reliable numerical data. 
The parallel tempering technique does not allow one to use
the equilibrium criteria used in Reference \cite{Bhatt_and_Young88},
in which two methods of calculating the disconnected spin glass susceptibility 
approach the equilibrium value from different directions. In the case of spin
glasses with Gaussian random bonds, a different technique was used
to determine the Monte Carlo time needed for equilibrated results
\cite{Katzgraber_Palassinia_Young}. This criterion cannot be used in the RFIM 
problem, but another one, which can be derived in a similar fashion, can be
used for Gaussian distributed random fields.
In the spirit of reference \onlinecite{Katzgraber_Palassinia_Young}, it
is straightforward to show that
\begin{equation}
q_{EA}=1-\frac{T}{N \Delta^2}\sum_i [h_i \langle S_i\rangle_T]_{av}\,\,\,,
\label{q_EA}
\end{equation}
where $q_{EA}\equiv (1/N)\sum_i[\langle S_i\rangle_T^2]_{av}$ 
is the Edwards-Anderson order
parameter, and $\langle \cdots\rangle_T$ and $[\cdots]_{av}$ 
indicate thermal and disorder average respectively. 
One can calculate $q_{EA}$ numerically using the overlap of two 
uncorrelated replicas ($a$ and $b$) at the same temperature, 
\begin{equation}
q_{EA}^{(1)}\equiv \frac{1}{N t_0}\sum_i \sum_{t=1}^{t_0} 
S_i^{(a)}(t)S_i^{(b)}(t)\,\,
\end{equation}
or by using expression (\ref{q_EA}) to obtain what we will call
$q_{EA}^{(2)}$. These two quantities approach the thermodynamic
equilibrium value $q_{EA}$ from opposite directions as $t_0$, the
number of Monte Carlo steps used, increases. Hence one can use
their convergence to a common value as a criterion for the equilibration
time.

We have not found an expression analogous to (\ref{q_EA}) for
the case of a bi-modal distribution of fields.
We note however that the time evolution of $q_{EA}^{(1)}(t_0)$ for 
the bimodal case follows that for the Gaussian distribution 
(using the same parameters) very closely. 
Therefore, we can take the equilibrating value of $t_0$  
obtained for the Gaussian distribution, and use     
it for the bi-modal case as a minimum 
Monte Carlo time needed to reach equilibrium for each   
disorder configuration. We retain the bi-modal distribution
for the bulk of our MC studies as it allows for a faster algorithm.

\begin{figure}
\epsfxsize=3.3in
\centerline{\epsffile{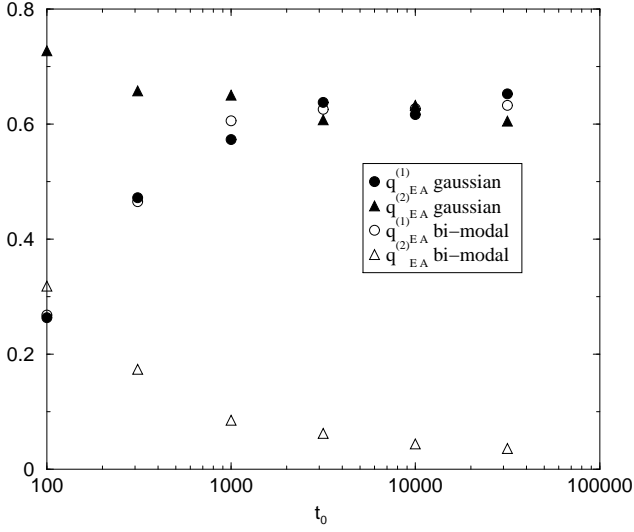}}
\caption{$q_{EA}^{(1)}$ and $q^{(2)}_{EA}$ vs.\ Monte Carlo time $t_0$ for
$T/J=2.2$, $\Delta/J=0.4$, and $L=5$. Here we show each quantity for the 
bi-modal (open symbols) and Gaussian (filled symbols) random field 
distribution.}
\label{equil_time}
\end{figure}

In Fig.\ \ref{equil_time}  we show the behavior of the two 
versions of $q_{EA}$, calculated with a Gaussian distribution
for $T/J=2.2$, $\Delta/J=0.4$, and $L=5$. 
We also show these two quantities for the bi-modal model.
It is clear that 
$q_{EA}^{(2)}$ is not equivalent to $q_{EA}^{(1)}$ in this case. 
However, as noted above, both $q_{EA}^{(1)}$ and $q_{EA}^{(2)}$ 
for the bi-modal model plateau
at the same point where $q_{EA}^{(1)}$ and $q_{EA}^{(2)}$ meet for the
Gaussian model. 

As criteria for equilibrium we then have several requirements. First, 
the equilibration time must be at least twice the time needed to establish 
convergence of $q_{EA}^{(1)}$ and $q_{EA}^{(2)}$ for the Gaussian case.
Second, we require that two distinct ways of computing 
$\chi_{SG}^{(1)}\equiv\frac{1}{N}\sum_{ij} \langle S_i S_j\rangle_T^2 $, 
one using two replicas at the same temperature 
and the other a direct Monte Carlo method, agree within statistical error
\cite{sccm}. 

\begin{figure}
\epsfxsize=3.3in
\centerline{\epsffile{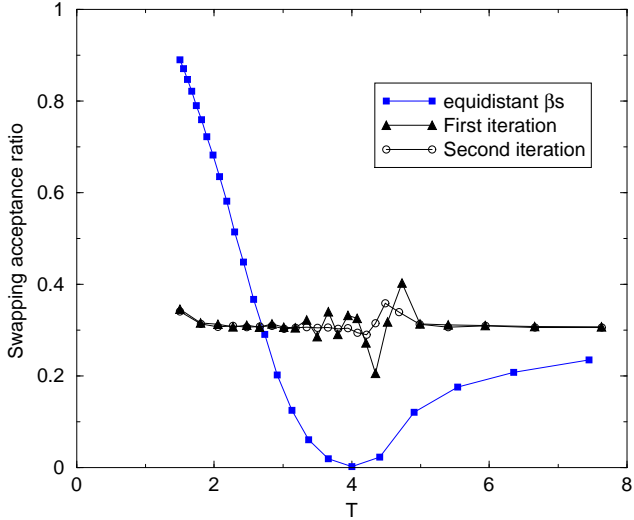}}
\caption{Swapping acceptance ratios among the configurations
at different temperatures at different stages
in the iteration defined by Eqs.\ (\ref{map}) and (\ref{map2}).
Here $L=11$, $T_{min}/J=1.5$,
and $\Delta/J=0.6$.}
\label{ex_betas}
\end{figure}

Our third criterion has to do with the exchange Monte Carlo method.
In order to ensure that equilibrium holds for all the various
temperatures simulated in parallel,
we require that the probability histogram for visiting the 
different temperatures is flat, and also
that the acceptance ratio for swapping among the configurations at 
different temperatures is at least $0.3$. 
In order to avoid the bottlenecks mentioned in 
reference \cite{Katzgraber_Palassinia_Young}
with regard to the acceptance ratios among the different 
temperatures, we follow a procedure introduced
by Hukushima \cite{Hukushima2}. Here we perform a quick Monte Carlo simulation
(only a few disorder realizations) with an equidistant  set of $\beta$'s 
to obtain an approximation for the average 
energy of the system. In order to obtain 
homogeneous acceptance ratios one performs the following mapping:
\begin{eqnarray}
\beta_n(t+1)=\frac{1}{2}[\beta_n(t)+g(\beta_n(t))]\,\,,
\label{map}
\end{eqnarray}
where
\begin{eqnarray}
g(\beta_n)=\frac{1}{E(\beta_{n-1})-E(\beta_{n+1})}\times
[\beta_{n-1} E(\beta_{n-1})\nonumber\\
-\beta_{n+1} E(\beta_{n+1})-E(\beta_n)(\beta_{n-1}-
\beta_{n+1})]\,.
\label{map2}
\end{eqnarray}
Here the values of $E(\beta_i)$ are obtained by extrapolation from the
ones obtained in the initial Monte Carlo runs. Using this procedure we obtain
a set of temperatures ($\beta$'s) that yield homogeneous acceptance
probabilities when a swap between two adjacent temperatures is attempted.
We can then, if necessary, increase the number of $\beta$'s and repeat
the procedure, until the uniform swapping probability reaches our
minimum goal of at least 0.3.
In Fig. \ref{ex_betas} we show the swapping acceptance ratios
after zero, one, and two iterations of
this procedure, for $L=11$, $T_{min}/J=1.5$, and $\Delta/J=0.6$. 
It is clear from the figure that the bottleneck at $T/J=4$ is removed
by the iterative procedure.

Our final criterion is to require that the maximum temperature
used in the parallel tempering process shows a `melted' $P(q)$; that is,
$P(q)$ at the smallest $\beta$ must have an approximate Gaussian shape
centered at or near $q=0$. 
This criterion is necessary to ensure that all free-energy
barriers have vanished at this $\beta$, so that the different
thermodynamic states in the low-temperature phase(s) will be
visited with a probability corresponding to their thermodynamic
weight.

\subsection{Numerical results}

As mentioned in Sec. I, there are indications 
from mean field studies
\cite{Almeida,Mezard_and_Monasson,Mezard_and_Young,Guagnelli_Marianari_Parisi} 
that a SG phase with RSB may exist
in an intermediate region of the phase diagram, which roughly
coincides with the irreversibility region found in dynamical calculations
\cite{Yoshizawa,Ro_Levin,Grest_Levin}.
One can see a hint of this by doing a 
Monte Carlo simulation, and computing the spectrum of $C_{ij}$,
for a single disorder realization
at a relatively small system size (in order for the calculation to 
be feasible) over the whole phase diagram.
We plot the
first two eigenvalues of $C_{ij}$ for a single disorder realization in 
Fig.\ \ref{Eval_T_vs_D}. In the FM and PM phases there is clearly a single
large eigenvalue $\lambda_1$, while in a region between the two phases
the first two eigenvalues are of the same order
of magnitude. This could be due to an intermediate phase
with RSB. On the other hand, such behavior could also be due to
the FM and PM eigenvectors (which represent two distinct forms of ordering)
trading places in the spectrum as
one moves from one phase to the other. 

\begin{figure}
\epsfxsize=3.375in
\centerline{\epsffile{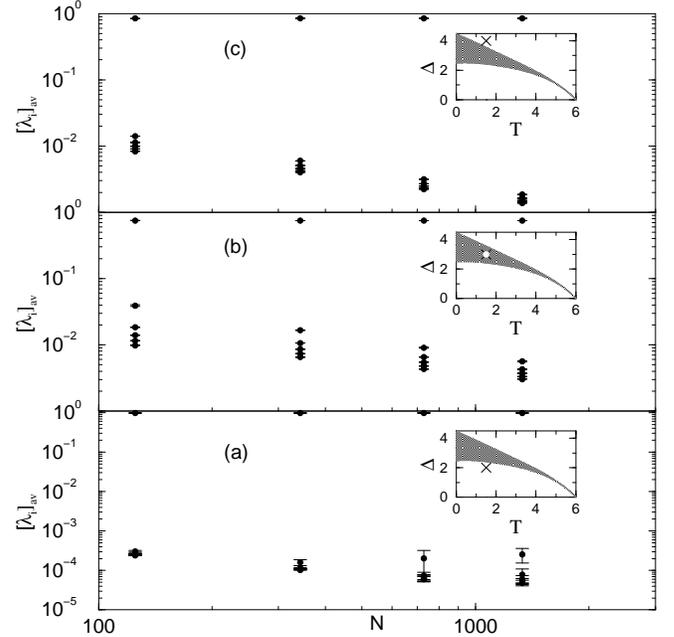}}
\caption{Scaling of the average of the first six eigenvalues $\lambda_i$
as a function of $N$ at $T/J=1.5$ and $\Delta/J=2.0$ (a),
$3.0$ (b), and $4.0$ (c). The location in the phase diagram obtained 
through dynamical Monte Carlo obtained from reference
3 is shown in the insert (here $J=1$).}
\label{phases}
\end{figure}

\begin{figure}
\epsfxsize=3.375in
\centerline{\epsffile{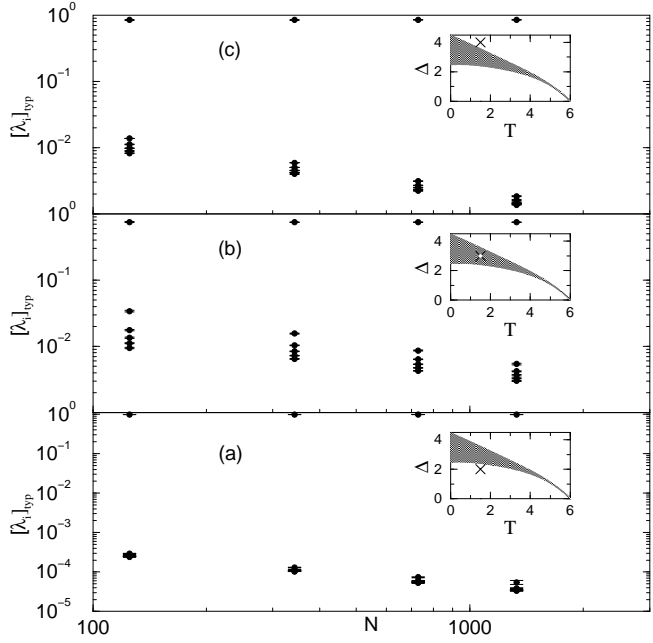}}
\caption{Scaling of the typical 
($[\lambda_i]_{\rm typ}\equiv\exp([\ln\lambda_i]_{\rm ave})$) 
of the first six eigenvalues 
as a function of $N$ at $T/J=1.5$ and $\Delta/J=2.0$ (a),
$3.0$ (b), and $4.0$ (c). The location in the phase diagram obtained 
through dynamical Monte Carlo obtained from reference
3 is shown in the insert (here $J=1$).
Note that, in this figure and in Figure \ref{phases}, the falloff
of $\lambda_2$ with $N$ becomes {\it weaker\/} at larger $N$ 
for $\Delta = 2.0$.}
\label{phases_typ}
\end{figure}

We can follow up these hints by studying larger systems, at a single
(low) temperature, over a range of fields $\Delta$ which spans the 
intermediate region seen in Fig.\ \ref{Eval_T_vs_D}.
Hence we have performed a
rather thorough Monte-Carlo scan of the field parameter $\Delta$
at a fixed, low temperature $T=1.5$.
For a preliminary coarse scan we studied the values
$\Delta/J=2.0,3.0$, and $4.0$.
The system sizes for each $\Delta$ are $L=5,7,9$, and 11,
with the number of disorder realizations being 2000, 2000, 850, and 650,
respectively.
Previous work \cite{Ro_Levin} has
yielded the dynamical phase diagram shown in the inserts of 
Figs. \ref{phases} and \ref{phases_typ}. 
Our coarse scan has been chosen to both bracket and sample the
irreversibility region, as shown in the inserts in
Figs. \ref{phases} and \ref{phases_typ}. In these figures we plot
(respectively) the average and typical values (where 
$[\lambda_i]_{\rm typ}\equiv
\exp([\ln \lambda_i]_{\rm ave})$) for the $\lambda_i/N$,
plotted against $N$ on a log scale. There is a clear power-law falloff
of $\lambda_2/N$ for $\Delta=3.0$ and $4.0$. The same is not quite true for
$\Delta = 2.0$. That is, for $\Delta = 2.0$ the falloff at small $N$ grows
weaker ($[\lambda]_{typ}$) or even ceases ($[\lambda]_{av}$) at larger
$N$. Therefore we performed a second, finer scan over the values
$\Delta = 1.8$, 2.2, 2.3, 2.4, and 2.6, with the same number of
of disorder realizations as given previously. 

Results from this scan ($[\lambda_i]_{\rm ave}$ only) are
plotted in Fig.\ \ref{finescan}. The hints of possible RSB seen in
Fig.\ \ref{Eval_T_vs_D} are considerably stronger here, particularly
at $\Delta = 2.4$. $\Delta = 1.8$ is clearly in the FM phase.
Also, it seems very likely that the falloff of $\lambda_2/N$
at $\Delta = 2.6$ will persist at larger $N$. Hence, we may take
$\Delta = 2.6$ to lie in the PM phase. The question is then,
have any of the other plots of $\lambda_2/N$ in Fig. \ref{finescan}
reached asymptotic behavior?

It is in fact impossible to answer this question from examination of
Fig.\ \ref{finescan} alone. One main problem is the rise in 
$\lambda_2/N$ that occurs in many of the plots at small $N$. 
This obvious finite-size effect is likely due to the fact
that we are not far from one (or two) phase boundaries. 
Such a rise must be followed by an asymptotic behavior which is either 
flat or falling; and, as is clear from the $\Delta = 2.6$ plot,
the latter can occur. For these reasons the flat region at
$\Delta=2.4$ gives no unambiguous conclusion: it may be just a local
maximum in $N$, or it may be the large-$N$ behavior. And we are
equally unsure of the large-$N$ behavior for $\Delta = 2.0$--$2.3$.
Taken as a whole, the six plots in Fig.\ \ref{finescan} can be viewed 
either as an RSB phase developing between two other phases, or as
the finite-size effects of a phase transition between two
competing order parameters.

\begin{figure}
\epsfxsize=3.15in
\centerline{\epsffile{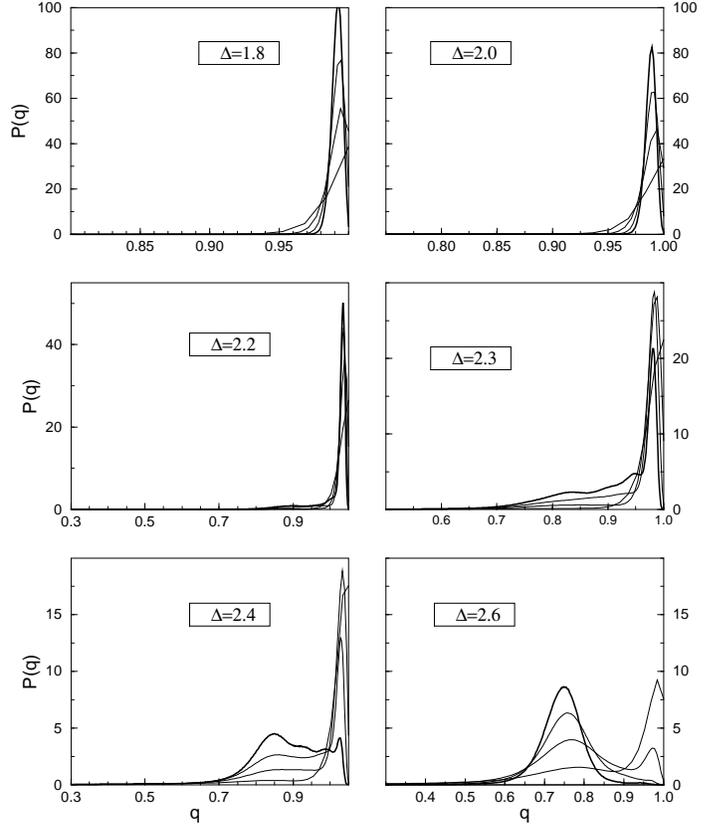}}
\caption{The overlap probability distribution $P(q)$, plotted
for various $\Delta$ (as indicated) and system size. The heavy curves
correspond to the largest size, $L = 11$. Curves (light lines)
for $L = 5$, 7, and 9
may be located by noting that, almost everywhere in $q$, the curves
monotonically approach the $L=11$ curve. Note that, for clarity, almost
every plot uses a different range on each axis.}
\label{Pofq}
\end{figure}

We find that this ambiguity is considerably diminished by a study of the
equilibrium $P(q)$ distributions for the various $\Delta$ values. 
Figure \ref{Pofq} shows the full $P(q)$ distribution for the same
$\Delta$ values seen in Fig.\ \ref{finescan}. The overall impression is
clear: two different kinds of order, represented by peaks at different
values of the overlap $q$, are competing (via fluctuations) in
the intermediate values of $\Delta$. At the lower end of the scan,
$\Delta=1.8$, the FM phase gives a peak at $q$ very close to one.
There is a much smaller peak at $q$ near $-1$ (not shown in the Figure)
which comes from fluctuations into the
``wrong'' spin ordering, i.e., that which is not favored by
the net random field [which is, again, of $O(N^{1/2})$].

At the other extreme ($\Delta=2.6$) it is equally clear that FM
fluctuations (present at small $N$) disappear at larger $N$, leaving a clear 
peak in $P(q)$ at the (lower) $q$ value appropriate to the PM phase. 
The dying off of these ferromagnetic fluctuations is well correlated
with the decay of $\lambda_2/N$ in Fig.\ \ref{finescan}.

\begin{figure}
\epsfxsize=3.375in
\centerline{\epsffile{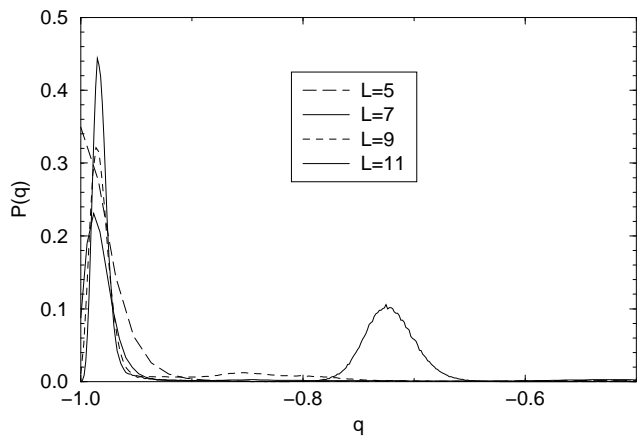}}
\caption{$P(q)$ for $\Delta = 2.2$, $-1 \le q \le -0.5$.
At the largest $L$ ($L = 11$)
we see two small (compare the vertical scale with Fig.\ \ref{Pofq})
but distinct peaks, which we ascribe to the overlap of ``wrong-sign"
FM fluctuations with ``right-sign" FM ordering (leftmost peak), and of
``wrong" FM fluctuations with fluctuations into
the paramagnetic phase (peak at $q \sim -0.73$).}
\label{Pofqnegq} 
\end{figure}

This interpretation is readily extended to the remaining 
$\Delta$ values. For instance, at $\Delta=2.2$ in 
Fig.\ \ref{Pofq} there are clearly PM fluctuations
giving a very broad peak at $q \sim 0.9$. This is even more clear 
if we expand the negative-$q$ portion of the $P(q)$ plot for the same
$\Delta$ (Figure \ref{Pofqnegq}). Here we see a small peak
near $q=-1$ showing overlap between ``wrong" and ``right"
FM states, and an even smaller peak which we ascribe to
overlap between ``wrong" FM fluctuations and PM fluctuations.
Very similar behavior is seen in $P(q)$ for $\Delta = 2.3$---both
at positive and negative $q$. The main difference is that the PM
fluctuations are stronger than they are at $\Delta = 2.2$.
Hence we conclude that both of these $\Delta$ values lie in the
FM phase for the RFIM, and that the growth of $\lambda_2/N$ with
increasing $N$ for these $\Delta$ values
is purely due to finite size, plus nearness to a phase boundary.

Finally, we see no justification for postulating a distinct
thermodynamic phase based on the behavior at the remaining 
$\Delta$ value ($\Delta = 2.4$).
The behavior both of its eigenvalue spectrum, and of its
$P(q)$ distribution, is clearly intermediate to the behaviors
at lower and higher $\Delta$. Furthermore, its $P(q)$
distribution, viewed on its own, gives no convincing sign
of converging towards a distribution appropriate for a spin glass.
In fact, the trend with $N$ (Fig.\ \ref{Pofq})
shows that the PM peak is steadily
growing with $N$, at the expense of the FM peak. This is, again,
a picture of competition between two forms of order (each
rather strong) near a phase transition,
rather than a distinct phase. Furthermore,
if one plots the total weight at negative $q$ [that is, the integral
$I_- = \int_{-1}^{0} dqP(q)$] against $N$ for the various $\Delta$
values, one finds very similar behavior for $\Delta = 2.6$ and $2.4$:
$I_-$ falls very fast, reflecting the fact that the ``wrong" FM 
fluctuations vanish rapidly with increasing system size. For all
of these reasons, we tentatively assign $\Delta = 2.4$ to the PM phase.

Figs.\ \ref{finescan}, \ref{Pofq}, and \ref{Pofqnegq}, viewed together,
give, we believe, a rather clear picture. There are two phases---FM
and PM---of the bimodal RFIM at $T=1.5 J$. The transition between these
two phases occurs (probably) between $\Delta = 2.3$ and $2.4$.

There is a further conclusion which seems obvious from the above
considerations, and from Fig.\ \ref{Pofq}. The location (in $q$)
of the peak in $q$, for the FM phase, shows no sign of going to zero as
the transition is approached from below. Hence, either the magnetization
is discontinuous at the phase boundary 
\cite{Hartmann,Angles,Swift,Machta,RiegerYoung,YoungNau} 
or it is continuous, but with an extraordinarily small exponent
$\beta$ \cite{Falicov,Ogielski}, such that it is `practically' discontinuous.
Obviously, our numerical studies cannot distinguish these two
possibilities. However, this aspect of the behavior of $P(q)$
at $T=1.5$---that is, the abrupt vanishing of the magnetization at
the transition---is completely consistent with the behavior seen
at higher temperatures (closer to the pure critical point)
and at zero temperature. This consistency, we believe, adds 
further support to our claim that there is no new, third phase
in the part of the phase diagram studied here.

\section{Conclusion}
\begin{figure}
\epsfxsize=3.375in
\centerline{\epsffile{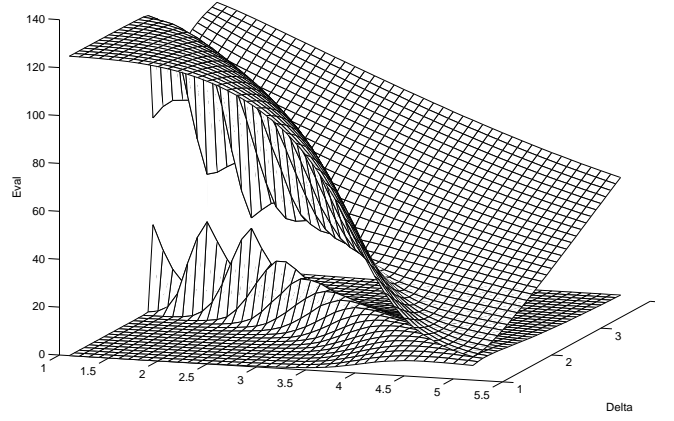}}
\caption{The first (upper sheet) and second (lower sheet) eigenvalue 
of the spin-spin correlation matrix for a single disorder realization.
The x-axis is temperature (label not indicated). Here $N=125$. The FM phase
lies in the front left corner (small $T$ and $\Delta$), 
where $\lambda_2 << \lambda_1$. The PM
phase lies on the other side of the ``ridge" in $\lambda_2$; in the PM phase it
is again true that $\lambda_2 << \lambda_1$. The ridge in $\lambda_2$,
with the corresponding ``trough" in $\lambda_1$, may indicate a distinct
thermodynamic phase; or it may simply indicate the effects of the FM/PM phase 
transition.}
\label{Eval_T_vs_D}
\end{figure}

\begin{figure}
\epsfxsize=3.15in
\centerline{\epsffile{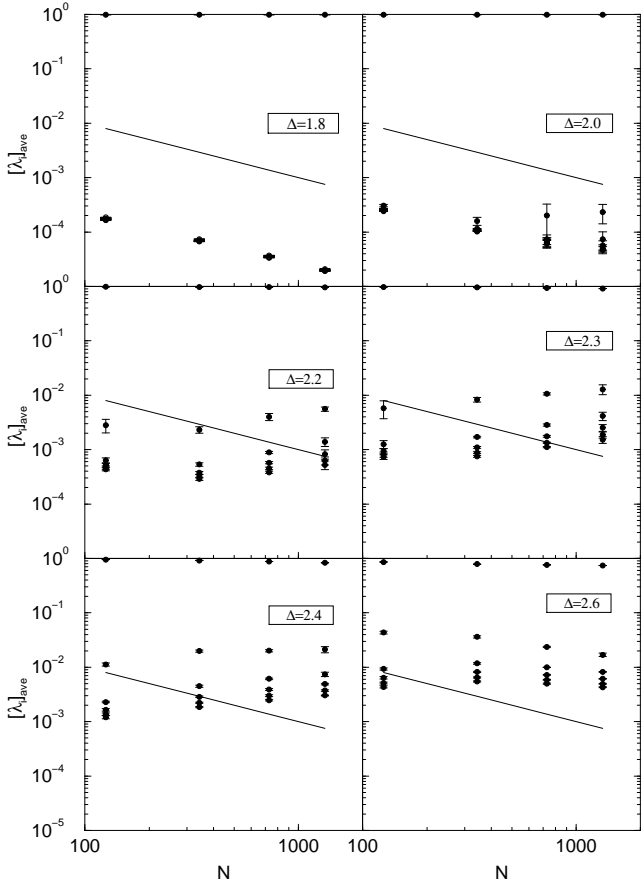}}
\caption{Scaling of the first six eigenvalues (disorder averaged)
at $T=1.5$, with $\Delta$ values as shown. The straight line, $y = 1/N$,
is included as a reference for the eye.}
\label{finescan}
\end{figure}

The above results and reasoning have presented a rather clear picture:
there is no spin-glass phase, and no replica symmetry breaking,
for the bimodal random-field Ising model
in the low-temperature region that we have examined. 
One of the tests that we have applied was to study the scaling with $N$
of the spectrum of eigenvalues of the spin-spin correlation function
$C_{ij}$. We have used this technique previously \cite{scm,sccm}
to give unambiguous confirmation of RSB in a case (the 
Sherrington-Kirkpatrick problem in infinite dimension) where it is
strongly believed to occur, and also to give strong evidence
against RSB for a finite-dimensional version of the same problem
(for which there is still controversy).
Here we have sought evidence for RSB in a different
problem, the RFIM---which may be mapped to the diluted 
antiferromagnet in an external field, a system which
may be studied experimentally. We have found
that the eigenvalue spectrum gives no sign at all of RSB
over most of the range of random field strength $\Delta$
that we have studied. There is a small range of $\Delta$ for
which the behavior of this spectrum is ambiguous.
However examination of the overlap distribution $P(q)$ removes
the ambiguity, giving instead clear evidence for a single phase transition
between the FM and PM phases, with no intervening equilibrium phase.
Hence we view the behavior of the eigenvalue spectrum
(Figs.\ \ref{Eval_T_vs_D} and \ref{finescan}) in this
intermediate region as purely the finite-size effects
of closeness to a phase boundary. In particular, the failure
of $\lambda_2/N$ to decay for this range of fields $\Delta$, 
even out to $N = 1331$ spins, stems from
the persistence of fluctuations between two adjacent phases, each with
a large order parameter. Hence we speculate that the
eigenvalues would behave less ambiguously---i.e., they would
display a reduced tendency to mimic RSB---if the transition were
continuous in a conventional sense,
such that the order parameters on either side vanished more smoothly.

It is possible (although, we feel, very unlikely) that studies
at larger system sizes may reveal these conclusions to be wrong,
confirming instead a spin-glass phase confined to a narrow range of 
$\Delta$ at low $T$.
A more promising direction for future work might be to
examine the RFIM with a Gaussian distribution of fields,
since it remains possible that the two problems have differing
phase diagrams. Also, one might argue that a SG phase is
suppressed by the strength of the order parameters on each
side of the transition. By this argument, one should look
at higher $T$, closer to the pure critical point. However,
both nonequilibrium \cite{Yoshizawa,Ro_Levin,Grest_Levin}
and equilibrium \cite{Almeida} mean-field arguments 
point to the low-temperature region as being most likely to
support a spin glass phase. This is why we looked at low $T$;
and our results rather clearly imply only two phases there.

The authors acknowledge helpful discussions with A.\ Aharony,
H. Castillo, and E. Fradkin. GSC thanks the Theory Group,
Physics Institute, University of Oslo, for hospitality.
This work was supported by the National Science
Foundation under Grant DMR-9820816 and the
Welch Foundation.

\end{document}